\documentclass[%
 reprint,
 amsmath,amssymb,
 aps,
 pre
]{revtex4-2}

\usepackage{graphicx}   
\usepackage{dcolumn}    
\usepackage{bm}         

\usepackage{tikz}

\usepackage[colorlinks=true,linkcolor=blue,urlcolor=blue,citecolor=blue]{hyperref}
\usepackage{xcolor}
\usepackage[normalem]{ulem}
\newcommand{\stkout}[1]{\ifmmode\text{\sout{\ensuremath{#1}}}\else\sout{#1}\fi}

\newcommand{\kT}{k_{\rm B}T}
\newcommand{\md}{{\rm d}}
\newcommand{\mc}[1]{\mathcal{#1}}

\newcommand{\V}[1]{V_{#1}}
\newcommand{\W}[2]{R_{#1}^{#2}}
\newcommand{\quantity}{transduced additional free energy rate }
\newcommand{\TAFER}{\dot{F}_{Y\to X}^{\rm add}}

\newcommand{\expow}{\beta\dot{W}_{Y\to X}^{\rm ex}}
\newcommand{\revpow}{\beta\dot{W}_{Y\to X}^{\rm rev}}

\begin{document}

\preprint{APS/123-QED}

\title{Free-energy transduction within autonomous systems}

\author{Steven J.\ Large}
\email{slarge@sfu.ca}
\author{Jannik Ehrich}
\author{David A.\ Sivak}
\email{dsivak@sfu.ca}
\affiliation{Department of Physics, Simon Fraser University, Burnaby, BC, V5A 1S6 Canada}

\date{\today}

\begin{abstract}
The excess work required to drive a stochastic system out of thermodynamic equilibrium through a time-dependent external perturbation is directly related to the amount of entropy produced during the driving process, allowing excess work and entropy production to be used interchangeably to quantify dissipation. Given the common intuition of biological molecular machines as internally communicating work between components, it is tempting to extend this correspondence to the driving of one component of an autonomous system by another; however, no such relation between the internal excess work and entropy production exists. Here we introduce the `transduced additional free-energy rate' between strongly coupled subsystems of an autonomous system, 
which 
is analogous to the excess power in systems driven by an external control parameter that receives no feedback from the system. We prove that this is a relevant measure of dissipation---in that it equals the steady-state entropy production rate due to the downstream subsystem---and demonstrate its advantages with a simple model system.
\end{abstract}

\maketitle

\section{\label{sec:introduction}Introduction}

Over the past 
25 
years, the development of stochastic thermodynamics has generalized the classical laws of macroscopic thermodynamics to strongly fluctuating microscopic systems~\cite{jarzynski_2011,seifert_2012,seifert_2019}. The theoretical framework of stochastic thermodynamics provides a method to assign physical quantities--such as work, heat, and entropy--to fluctuating systems in contact with thermodynamic reservoirs, even far from equilibrium. These physical quantities can be identified along a single stochastic trajectory, or at the level of probability distributions, allowing for a diverse set of methods to understand the physics of thermodynamic systems across all scales~\cite{van_den_broeck_2014}.

In parallel with the formulation of stochastic thermodynamics, a variety of experimental techniques have been developed that directly test and verify its predictions through passive observation, or direct manipulation of an external control parameter~\cite{ciliberto_2017}. For instance, measurements of heat and work fluctuations have been made in microscopic systems like an AFM cantilever~\cite{gomez_solano_2010}, a torsion pendulum~\cite{douarche_2006}, or an electrical resistor~\cite{garnier_2005}. Additionally, the widespread utilization of optical tweezers for single-molecule force spectroscopy has allowed for the direct verification of theoretical predictions in biological systems, such as a DNA hairpin~\cite{tafoya_2019,collin_2005}.  

Within this experimental paradigm, where a stochastic system is manipulated using an external device, there is a convenient correspondence between the excess work done on the system and the total entropy production~\cite{esposito_2010_PRL}. Specifically, the entropy production fluctuation theorem~\cite{seifert_2005} and the Crooks fluctuation theorem~\cite{crooks_1999} are equivalent for a stochastic system---initially at thermodynamic equilibrium---driven by an external control parameter that receives no feedback from the system~\cite{esposito_2010_PRL}. Here, the excess work and entropy production can be used interchangeably when referring to dissipation. This correspondence is a powerful tool, granting the freedom to quantify dissipation either through work done in excess of free-energy changes, or system entropy changes not compensated by heat flows. However, the same such correspondence does not hold in more complex coupled systems~\cite{horowitz_2014}.

While initial experimental tests of stochastic thermodynamics focused primarily on verifying the so-called fluctuation theorems~\cite{ciliberto_2017}, the use of stochastic thermodynamics in ever-more complex systems has become commonplace. In fact, a central motivation for the field of stochastic thermodynamics has been to better understand the complex molecular machines in biological systems~\cite{seifert_2012}. These nanoscale machines consist of strongly interacting protein complexes, which interconvert between different forms of energy to perform useful functions within the cell~\cite{brown_2020}. For instance, the rotary machine ${\rm F}_{\rm o}{\rm F}_1$ ATP synthase makes use of a proton gradient across the mitochondrial membrane to catalyze the production of the chemical energy currency ATP~\cite{toyabe_2011,noji_1997}, while the transport motor kinesin utilizes the energy stored in ATP to directionally transport cargoes throughout the cell~\cite{valentine_2006}.

Entropy production represents a fundamental loss in the operation of a machine: low-entropy states of a system (all gas molecules confined to one half of a cylinder, a fully extended polymer) or of thermal baths (baths at different temperatures) can be harnessed to do useful work (push a piston during expansion, exert a force during compaction, drive a heat engine during heat flow), whereas high-entropy states have reduced capacity to do so~\cite{schroeder}. Thus in biophysical molecular machines, entropy production serves as an important performance measure.

Although such machines operate autonomously, there are many natural comparisons between controlled systems and molecular machines. For instance, it is conceptually straightforward to view the chemical hydrolysis of ATP as driving the processive motion of the molecular motor kinesin~\cite{wagoner_2019}, or the ${\rm F}_{\rm o}$ component of ATP synthase as mechanically driving the rotation of a central crankshaft, inducing the ${\rm F}_{\rm 1}$ component to catalyze the production of ATP~\cite{toyabe_2011}. Thus there is a natural appeal to quantifying the work (and the excess work) done by one component of an autonomous biomolecular machine on another. Indeed, such internal energy flows have been used to aid in the identification of reaction coordinates in biomolecular dynamics~\cite{li_2016}. 

Given the biophysical importance of entropy production, and the mathematical constraints imposed on it by fluctuation theorems and second-law-like inequalities, it is tempting to use the correspondence between excess work and entropy production in systems driven by an external control parameter to aid in the conceptual and quantitative understanding of biomolecular machines~\cite{bryant_2020,large_2018,machta_2015}. However, while it is still possible to define excess work internal to such systems as an energy flow, it has no direct relationship to the entropy production.

In this article, we investigate excess work and entropy production in strongly coupled autonomous systems, and present an alternative measure of dissipation---the \emph{transduced additional free-energy rate}---which plays the same thermodynamic role in autonomous systems as excess power does in externally driven systems. Specifically, the steady-state \quantity (differing from the excess power by an information rate that captures the effects of inter-system coupling) equals an entropy production rate. We then illustrate our results in a simple model of a cyclic mechanochemical motor, where a mechanical system is driven via its strong coupling to a stochastic nonequilibrium chemical reaction.

\section{\label{sec:excess-power}
Strongly coupled multi-component systems}

Throughout this article, we consider a bipartite system described by two coordinates $X$ and $Y$, with states $x$ and $y$, respectively. The system's joint dynamics are described by a discrete-state master equation~\cite{hartich_2014,horowitz_2014}
\begin{equation}
    \md_t p_{xy} = \sum_{x',y'}\W{yy'}{xx'} p_{x'y'} \ , \label{master-equation}
\end{equation}
where $p_{xy}$ is the joint probability of state $(x,y)$, $\md_t~\equiv~\md/\md t$ is the time derivative, and $\W{yy'}{xx'}$ is the transition rate matrix element quantifying the rate of the transition $(x',y') \to (x,y)$. 

The assumption of bipartite dynamics restricts the entries of the transition rate matrix~\cite{hartich_2014,horowitz_2014}:
\begin{equation}
 	\W{yy'}{xx'} = \begin{cases}
		\qquad\quad \W{yy'}{x} \quad&{\rm if}\quad y\neq y'\,,\, x=x' \\
		\qquad\quad \W{y}{xx'} \quad&{\rm if}\quad y=y'\,,\,  x\neq x' \\
		\qquad\quad\; 0 \quad& {\rm if}\quad y\neq y'\,,\,  x\neq x' \\
		{-\sum\limits_{x \neq x'}\W{y}{xx'}}-\sum\limits_{y \neq y'}\W{yy'}{x} & {\rm if}\quad y=y'\,,\,  x= x'
 	\end{cases}
 	\ .
\end{equation}

Thus, for the ensemble average $A_{XY} \equiv \sum_{x,y}p_{xy} a_{xy}$ of the fluctuating quantity $a_{xy}$ that depends on the joint state $(x,y)$, the time rate of change $\md_t A_{XY} = \sum_{x,x',y,y'} \W{yy'}{xx'}p_{x'y'}(a_{xy} - a_{x'y'})$ can be decomposed into contributions due to the individual dynamics of each subsystem:
\begin{subequations}
\begin{align}
    \md_t A_{XY} &= \sum_{x,x',y} \W{y}{xx'}p_{x'y}(a_{xy} - a_{x'y}) \nonumber \\
    &\quad + \sum_{x,y,y'}\W{yy'}{x}p_{xy'}(a_{xy} - a_{xy'})
    \label{time-deriv-decomp-a} \\
    &= \dot{A}^X + \dot{A}^Y \label{time-deriv-decomp-b} \ .
\end{align} \label{time-deriv-decomp}%
\end{subequations}
The overdot indicates that---unlike the time derivative $\md_t A_{XY}$---the individual rates ($\dot{A}^X$ and $\dot{A}^Y$) cannot be written as the time derivative of a function. Upper-case letters indicate ensemble-averaged quantities.

\subsection{Entropy production}
The entropy production rate (for unitless entropy) of the joint system
is~\cite{horowitz_2014}
\begin{equation}
    \dot{\Sigma} = \sum_{x,y,x',y'}\W{yy'}{xx'}p_{x'y'}\ln\frac{\W{yy'}{xx'} p_{x'y'}}{\W{y'y}{x'x} p_{xy}} \label{total-entropy-production} \geq 0 \ .
\end{equation}

The total entropy production can be conveniently split into separate contributions arising from the respective transitions among subsystems $Y$ and $X$:
\begin{subequations}
\begin{align}
    \dot{\Sigma} &= \dot{\Sigma}^Y + \dot{\Sigma}^{\rm X} \\
	&= \sum_{x,y,y'}\W{yy'}{x} p_{xy'}\ln\frac{\W{yy'}{x} p_{xy'}}{\W{y'y}{x}p_{xy}} \\
	&\quad + \sum_{x,x',y}\W{y}{xx'} p_{x'y}\ln\frac{\W{y}{xx'} p_{x'y}}{\W{y}{x'x} p_{xy}} \ . \nonumber
\end{align} \label{entropy-split}%
\end{subequations}
Each RHS entropy production rate obeys a second-law-like inequality, 
\begin{subequations}
\label{modified-second-law}
\begin{align}
    \dot{\Sigma}^X &= \md_t S_X + \dot{S}_{\rm e}^X - \dot{I}^X \geq 0
	\label{modified-second-law-x} \\
	\dot{\Sigma}^Y &= \md_t S_Y + \dot{S}_{\rm e}^Y - \dot{I}^Y \geq 0 \ .
	\label{modified-second-law-y}
\end{align}
\end{subequations}
$\md_t S_X$ ($\md_t S_Y$) is the rate of change of the entropy of subsystem $X$ ($Y$). $\dot{S}_{\rm e}^X$ ($\dot{S}_{\rm e}^Y$) is the rate of entropy flow from the system to the heat reservoir due to $X$ ($Y$) dynamics, which for a thermodynamic system coupled to a heat bath (all systems considered here) equals $-\beta \dot{\mc{Q}}^X$ ($-\beta \dot{\mc{Q}}^Y$), the negative rate of heat flow into the system---and hence entropy increase in the heat reservoir---due to $X$ ($Y$) dynamics. (By convention, positive work and heat correspond to energy flow into the system.)

The information rates are
\begin{subequations}
\begin{align}
	\dot{I}^X &= \sum_{x,x',y}\W{y}{xx'}p_{x'y}\ln\frac{p_{y|x}}{p_{y|x'}} \label{info-rate-x} \\
	\dot{I}^Y &= \sum_{x,y,y'}\W{yy'}{x}p_{xy'}\ln\frac{p_{x|y}}{p_{x|y'}} \ , \label{info-rate-y}
\end{align}\label{info-rate}%
\end{subequations}
for conditional probability $p_{x|y} \equiv p_{xy}/\sum_x p_{xy}$ of state $x$ given state $y$. $\dot{I}^X$ ($\dot{I}^Y$) represents the rate of change of mutual information between subsystems $X$ and $Y$ due to transitions in $X$ ($Y$)~\cite{hartich_2014,horowitz_2014}. Thus, a positive (negative) information rate $\dot{I}^X$ implies that, on average, dynamics of $X$ increase (decrease) the mutual information between the two subsystems.

At steady state, the joint-system entropy is unchanging ($\md_t S_{XY} = 0$), as are the entropies of each subsystem ($\md_t S_X = \md_t S_Y = 0$). Thus, the combined effect of $X$ and $Y$ dynamics leaves the mutual information unchanged, so the information rates are opposite ($\dot{I}^X = -\dot{I}^Y$) and cancel when summing the entropy production rates $\dot{\Sigma}^X$ and $\dot{\Sigma}^Y$ of each subsystem [Eq.~\eqref{modified-second-law}]~\cite{horowitz_2014}. Therefore, the total steady-state entropy production is the sum of each subsystem's heat flows:
\begin{equation}
	\dot{\Sigma} = -\beta\dot{\mc{Q}}^Y - \beta\dot{\mc{Q}}^X \ .
	\label{entropy-production-reservoir}
\end{equation}
However, unlike the entropy productions of each subsystem in Eq.~\eqref{modified-second-law}, neither RHS term is lower bounded by zero.

\subsection{Excess work}
To quantify the energy flow in such a system, we now treat subsystem $Y$ as a work source for subsystem $X$, so that the average rate of work (the average power) done by $Y$ on $X$ is
\begin{equation}
	\beta\dot{W}_{Y\to X} = \sum_{x,y,y'}\beta\left( \epsilon_{xy} - \epsilon_{xy'} \right)\W{yy'}{x} p_{xy'} \ , \label{autonomous-power}
\end{equation}
for energy $\epsilon_{xy}$ of state $(x,y)$. Throughout this article we exclusively deal with average power and average work, so for the remainder we omit explicit mention of averages.  

For fixed $y$, the conditional equilibrium distribution of $X$ is $\pi_{x|y} = \exp(-\beta\epsilon_{xy} + \beta F_{X|y})$, for conditional equilibrium free energy $F_{X|y} \equiv - \kT \ln\sum_x \exp(-\beta\epsilon_{xy})$. Thus, the energy can be expressed in terms of the conditional equilibrium distribution and free energy as
\begin{equation}
\epsilon_{xy} = -\kT \ln\pi_{x|y} + F_{X|y} \ . \label{boltzmann-conditional}
\end{equation}

The power done on $X$ is the sum of the reversible power and the excess power,
\begin{subequations}
\begin{align} 
	\beta\dot{W}_{Y\to X} &= \revpow + \expow \\
	&= \beta\sum_{x,y,y'}\W{yy'}{x} p_{xy'} (F_{X|y}-F_{X|y'}) \nonumber\\
	&\qquad\qquad\qquad+\sum_{x,y,y'} \W{yy'}{x} p_{xy'}\ln\frac{\pi_{x|y'}}{\pi_{x|y}} \ .
\end{align}
\end{subequations}
The excess power quantifies the rate of energy flow from subsystem $Y$ into subsystem $X$ which exceeds the rate of change of the conditional equilibrium free energy of $X$.

The excess power can be simplified by separating the $X$ and $Y$ summations, 
\begin{subequations}
\begin{align}
	\expow &= \sum_{y,y'}\V{yy'}p_{y'}\left[\sum_x \frac{\W{yy'}{x}p_{x|y'}}{\V{yy'}}\ln\frac{\pi_{x|y'}}{\pi_{x|y}} \right] \\
	&= \sum_{y,y'}\V{yy'}p_{y'} \beta W^{\rm ex}_{yy'} \ .
\end{align}\label{excess-power}%
\end{subequations} 
Here $p_y \equiv \sum_x p_{xy}$ is the marginal distribution of state $y$, and the $X$-summation is simply the excess work on $X$ during transition $y'\to y$. The coarse-grained rate 
\begin{equation}
    \V{yy'} \equiv \sum_x \W{yy'}{x} p_{x|y'} \label{coarse-grained-rate}
\end{equation}
is the rate of $y'\to y$, averaged over the conditional distribution of $X$~\cite{esposito_2012}. $\V{yy'}$ normalizes the switching-state distribution 
\begin{equation}
    p_{yy'}^{{\rm sw},x} \equiv \frac{\W{yy'}{x} p_{x|y'}}{\V{yy'}} \ , \label{switching-state-dist}
\end{equation}
the conditional distribution of $X$ during a $y'\to y$ transition.

The excess work per $y'\to y$ transition can be expressed using Eq.~\eqref{switching-state-dist} as the difference between the relative entropies of the switching-state distribution with the respective conditional equilibrium distributions after ($\pi_{x|y}$) and before ($\pi_{x|y'}$) the transition:
\begin{equation} 
    \beta W^{\rm ex}_{yy'} = D\left( p_{yy'}^{{\rm sw},x} \, || \, \pi_{x|y} \right) - D\left( p_{yy'}^{{\rm sw},x} \, || \, \pi_{x|y'} \right) \label{excess-work-relative-entropy} \ .
\end{equation}
The relative entropy is defined as $D(p_x||q_x) \equiv \sum_x p_x \ln(p_x / q_x)$~\cite{cover_thomas}.

\section{Classes of upstream dynamics}

\subsection{\label{subsec:deterministic-work}External control parameter} 
Insofar as it relates to entropy production, the excess power is a quantity of interest in many systems driven by an external control parameter. Many experimental manipulations of machines drive the system according to dynamics of an experimental apparatus that do not depend on the system response, i.e., with no feedback. This corresponds to the special case of the above framework where the $Y$ dynamics are independent of the current state of $X$. Here, we consider stochastic $Y$ dynamics~\cite{bryant_2020,large_2018,machta_2015}; however, in general they could alternatively be deterministic~\cite{large_2019,sivak_2012,aurell_2011,schmiedl_2007}.

Since such independent $Y$ dynamics ensure the conditional independence of the target state $y$ and the current mechanical state $x$ (conditioned on the source state $y'$), the data-processing inequality~\cite{cover_thomas} requires that the $Y$ dynamics reduce the mutual information between subsystems: $\dot{I}^Y \leq 0$. Mathematically, for independent $Y$ dynamics the information rate $\dot{I}^Y$ [Eq.~\eqref{info-rate-y}] can be written as a negative relative entropy, $\dot{I}^Y = -\sum_{y,y'}\V{yy'}\,p_{y'} D\left( p_{x|y'} || p_{x|y} \right) \leq 0$, and thus is necessarily non-positive. Therefore, at steady state (where $\dot{I} = 0$), $\dot{I}^X \geq 0$, and the form of the second law $\beta\dot{\mc{Q}}^X \leq 0$ holds for the heat flow $\beta\dot{\mc{Q}}^X$ due to $X$ dynamics~\footnote{This insight allows some intriguing interpretations in the context of the \emph{thermodynamics of sensing}, where a system $X$ collects information about an external and independent stochastic variable $Y$. Rearranging the second law with the information rate on the RHS of Eq.~\eqref{modified-second-law} yields a refined lower bound on the steady-state dissipation for the system in terms of the \emph{nostalgia}~\cite{still_2012,quenneville_2018} or \emph{learning rate}~\cite{barato_2014,brittain_2017}.}.

For independent $Y$ dynamics, the switching-state distribution is the conditional distribution $p_{x|y'}$ of $x$ given the source state $y'$, independent of the target state $y$, and the excess work for transition $y'\to y$ is
\begin{equation}
	\beta {W}^{\rm ex}_{yy'}
	= D\left( p_{x|y'} \, || \, \pi_{x|y}\right)  - D\left(p_{x|y'} \, || \,  \pi_{x|y'}\right) \label{excess-work-relative-entropy-determ} \ .
\end{equation}
In the timescale-separated limit~\cite{esposito_2012}, where the $X$ dynamics are much faster than the $Y$ dynamics, the conditional distribution over $X$ equilibrates between each $Y$ transition, so the second RHS term is zero, reducing Eq.~\eqref{excess-work-relative-entropy-determ} to the \emph{infinite-time excess work}~\cite{large_2019},
\begin{equation}
	\beta W^{\rm ex}_{yy'} = D\left(\pi_{x|y'}||\pi_{x|y}\right) \geq 0 \ . \label{excess-work-determ-tss} 
\end{equation} 
This excess work is non-negative for any transition, and thus the excess power [Eq.~\eqref{excess-power}] is positive, even for no net $Y$ flux, $\V{yy'} p_{y'} = \V{y'y} p_{y}$ (App.~\ref{app:autonomous-vs-external} gives details) \footnote{Reference~\cite{large_2018} found similar behavior, where the excess power (Eq.~(9) in Ref.~\cite{large_2018}) to drive a system through an ensemble of stochastic control protocols--independent of the system response--contains a term that is independent of the driving strength.}.

\subsection{\label{subsec:transduced-power}Thermodynamically complete system} 

In autonomous systems (such as molecular machines consisting of multiple strongly interacting components) not subject to temporal variation of an external control parameter, thermodynamic consistency requires that the entries of the transition rate matrix $\W{yy'}{xx'}$ satisfy local detailed balance~\cite{seifert_2019,van_den_broeck_2014,bergmann_1955},
\begin{equation}
    \ln\frac{\W{yy'}{xx'}}{\W{y'y}{x'x}} = -\beta\left(\Delta\epsilon_{yy'}^{xx'} + \Delta\mu_{yy'} \right)
	\label{local-detailed-balance} \ ,
\end{equation}
where $\Delta\epsilon_{yy'}^{xx'} + \Delta\mu_{yy'}$ is the change in thermodynamic potential during the transition $(x',y')\to (x,y)$, involving the change $\Delta\epsilon_{yy'}^{xx'} \equiv \epsilon_{xy}-\epsilon_{x'y'}$ in system energy during transitions in $X$ and $Y$ and the change $\Delta\mu_{yy'}$ in chemical potential during transitions of subsystem $Y$ (satisfying $\Delta\mu_{yy'}=-\Delta\mu_{y'y}$ and hence $\Delta\mu_{yy}=0$). Despite the particular form of thermodynamic potential implied by the RHS, the theoretical framework we present is more broadly applicable so long as the dynamics of subsystem $X$ are detailed balanced.

We call systems \emph{thermodynamically complete} when all rates satisfy local detailed balance [Eq.~\eqref{local-detailed-balance}]. Conversely, we call systems (such as the independent $Y$ dynamics in Sec.~\ref{subsec:deterministic-work}) \emph{thermodynamically incomplete} when the transition rates violate Eq.~\eqref{local-detailed-balance}, as some external influences are required to ensure thermodynamic consistency. Thus, thermodynamically complete systems are those which in the absence of driving relax to equilibrium, though with driving present (our case here) they need not.

In detailed-balanced dynamics---or any dynamics where subsystem $Y$ receives feedback from $X$---the excess work [Eq.~\eqref{excess-work-relative-entropy}] associated with a particular $Y$ transition is not lower bounded by zero, and can be negative. 

We present the \textit{\quantity} or TAFER (the name will become clear),
\begin{equation} 
    \beta\TAFER \equiv \expow + \dot{I}^Y \label{transduced-power} \ ,
\end{equation}
as a measure of dissipation between strongly coupled subsystems that is analogous to the excess power [Eq.~\eqref{excess-power}] in systems driven by an external control parameter. Appendix~\ref{app:trans-power} gives a detailed derivation.

Unlike the excess power, however, at steady state TAFER is lower bounded by zero: $\expow = -\beta\dot{\mc{Q}}^X$ (see App.~\ref{app:excess-power-entropy-prod}) and $\dot{I}^Y = -\dot{I}^X$, so Eq.~\eqref{transduced-power} coincides with Eq.~\eqref{modified-second-law-x}. In terms of the underlying probability distributions, 
\begin{equation}
	\beta\TAFER = \sum_{y,y'}\V{yy'} p_{y'}\left[ \sum_x p_{yy'}^{{\rm sw}, x}\ln\frac{\pi_{x|y'}p_{x|y}}{\pi_{x|y}p_{x|y'}} \right] \ . \label{transduced-power-explicit}
\end{equation}
TAFER modifies the excess power in Eq.~\eqref{excess-power} by the additional average of the log-ratio of nonequilibrium conditional distributions over the switching-state distribution $p_{yy'}^{{\rm sw}, x}$. This form clarifies that TAFER vanishes for all transitions $y'\to y$ in the timescale-separated limit, where $p_{x|y} \to \pi_{x|y}$.

TAFER can also be written as (see App.~\ref{app:trans-power-add-free-energy})
\begin{align}
  \beta\TAFER = \beta\dot{F}^{{\rm neq},Y}_{X|Y} - 
  \beta \, \mathrm{d}_t F_{X|Y}\ , 
\end{align}
where $\dot{F}^{{\rm neq},Y}_{X|Y}$ is the change of conditional nonequilibrium free energy~\cite{Esposito2011NonequilibriumFreeEnergy} due to the $Y$ dynamics, and $\mathrm{d}_t F_{X|Y}$ is the change in equilibrium free energy of $X$ given $Y$. Thus, the \quantity is indeed the rate of change due to the $Y$ dynamics of the additional free energy in $X$~\cite{Esposito2011NonequilibriumFreeEnergy} (above the equilibrium free energy)
\begin{align}
    F^{\rm add}_{X|Y} \equiv F^{\rm neq}_{X|Y} - F_{X|Y} \ .
\end{align}
This clarifies that TAFER quantifies how much the $Y$ dynamics contribute to $X$ being out of conditional equilibrium.

\section{\label{sec:model-system}Model system}
\begin{figure}
	\centering\includegraphics[width=\columnwidth]{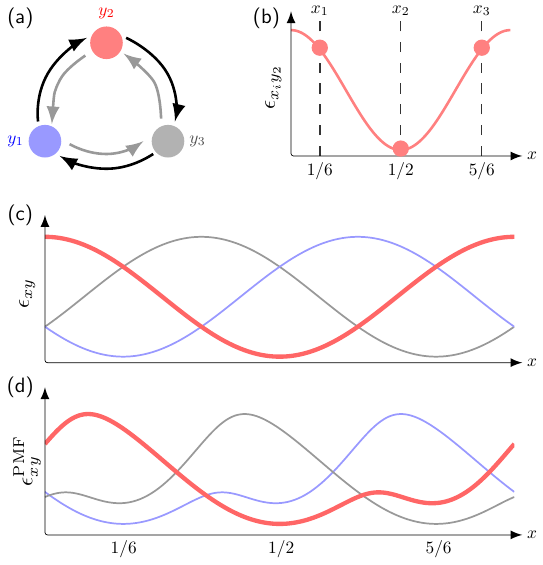}
	\caption{\label{fig:schematic}{\bf Schematic of the model mechanochemical system}. 
	(a) 
	Three-state 
	chemical reaction network representing $Y$ dynamics.
	(b) The imposed energy landscape on the mechanical coordinate $X$ (with equidistant discrete states) for $Y=y_2$ and $N=3$ mechanical states (Sec.~\ref{sec:expowNEEP}). 
	Each specific chemical state $y$ imposes on $X$ a periodic energy landscape, with either (c) a single minimum (Sec.~\ref{sec:expowNEEP}) or (d) two local minima (Sec.~\ref{sec:expowNeg}).
    }
\end{figure}

We now illustrate our theory in a minimal model of a mechanochemical molecular machine. Figure~\ref{fig:schematic} shows a schematic.

The chemical coordinate $Y$ evolves on a 
three-state 
cycle, ($y_1, y_2, y_3$) = (1/6, 1/2, 5/6), increasing in the clockwise direction, while the mechanical coordinate $X$ contains $N$ discrete states, ($x_1, x_2, \ldots, x_n$) where $x_k = (k-\tfrac{1}{2})/N$, evolving on a periodic energy landscape imposed by the current value of $Y$. Subsystems $X$ and $Y$ 
have transitions 
only 
between nearest-neighbor states, and obey periodic boundary conditions. The instantaneous transition rates for $X$ and $Y$---satisfying local detailed balance [Eq.~\eqref{local-detailed-balance}]---are
\begin{subequations}
\begin{align} 
	\W{yy'}{x} &= \Gamma_{\rm chem}\exp\left\{ -\tfrac{1}{2}\beta\left( \Delta\mu_{yy'} + \Delta\epsilon_{yy'}^{x} \right) \right\} \\
	\W{y}{xx'} &= \Gamma_{\rm mech}\exp\left\{ -\tfrac{1}{2}\beta\Delta\epsilon_y^{xx'} \right\} \ ,
\end{align}\label{simulation-rates}%
\end{subequations}
where $\Gamma_{\rm chem}$ and $\Gamma_{\rm mech}$ are kinetic prefactors for the chemical and mechanical rates, quantifying the \textit{bare rates} of each process in the absence of any differences $\Delta\epsilon_{yy'}^{xx'} \equiv \epsilon_{xy} - \epsilon_{x'y'}$ in state energies $\epsilon_{xy}$ or differences $\Delta\mu_{yy'}$ in chemical potentials~\cite{brown_2017}. Thus chemical transitions---indicated by changes in chemical potential---and energy changes fully determine the transition-rate asymmetries. Here, the thermodynamic potential is a function of the state of the system and the chemical reservoirs (the `super-system'), thus while a single cycle of the chemical ($Y$) subsystem ($y_1\to y_2\to y_3\to y_1$ in Fig.~1a) returns the chemical coordinate $Y$ to the same state, the chemical reservoirs are in a different state due to a net transfer of particles. However, since each transition in the $Y$ subsystem is associated with a known change in the reservoirs $\Delta\mu_{yy'}$, we can nevertheless specify a thermodynamically complete dynamics of the $XY$ system that is out of equilibrium without explicitly tracking the state of the chemical reservoirs~\cite{seifert_2019}.

We further assume identical chemical potential differences ($\Delta\mu_{y_{i+1},y_i} = -\Delta\mu_{y_i,y_{i+1}} = \Delta\mu < 0$, for $i+1$ taken modulo $3$) for each chemical transition, and (without loss of generality) that negative $\Delta\mu$ induces net clockwise rotation of the chemical coordinate. Physically, the chemical potential differences are generated by out-of-equilibrium concentrations of products and reactants, such as ATP and ADP for many molecular machines.

\subsection{Excess power does not equal entropy production\label{sec:expowNEEP}}

\begin{figure} 
    \centering\includegraphics[width=\columnwidth]{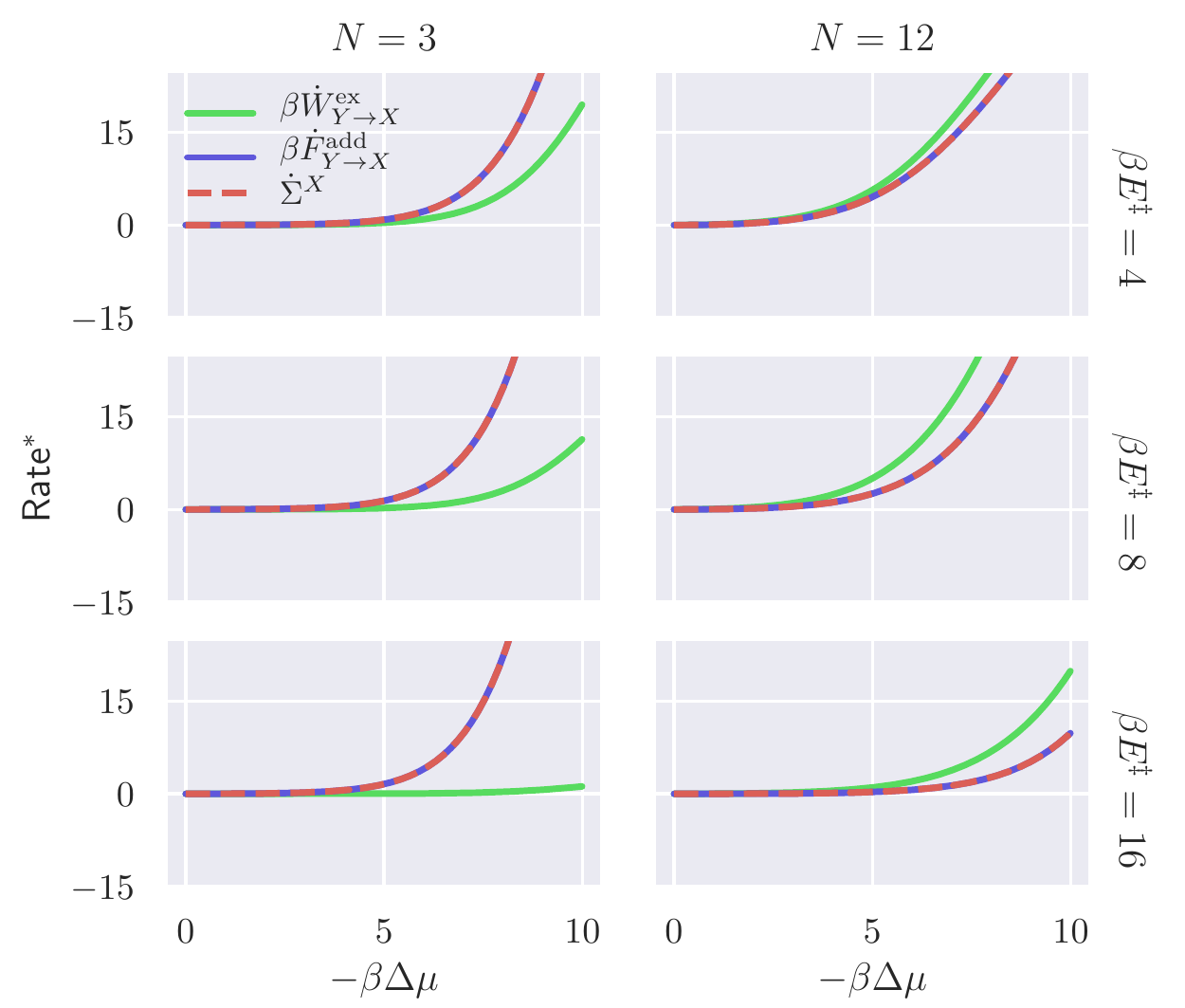}
    \caption{\label{fig:dissipation}{\bf At steady state, \quantity equals entropy production rate, but excess power need not.} 
    The excess power $\expow$ (green), \quantity $\beta\TAFER$ (blue), and entropy production rate $\dot{\Sigma}^X$ (red dashed) due to $X$ dynamics, each as a function of the chemical driving strength  $-\beta\Delta\mu$, for a range of barrier heights $\beta E^{\ddagger} = 4,8,16$ (rows) and numbers $N = 3,12$ of mechanical states (columns). For all panels, the system is at steady state for potential~\eqref{periodic-energy}, $\Gamma_{\rm mech}/\Gamma_{\rm chem} = 100$, and rates are nondimensionalized by the chemical bare rate: Rate$^* \, \equiv$ Rate/$\Gamma_{\rm chem}$.}
\end{figure}

First, we consider a periodic monostable potential,
\begin{equation}
    \epsilon_{xy} = \tfrac{1}{2}E^{\ddagger}\cos 2\pi\left(x - y \right) \label{periodic-energy} \ ,
\end{equation}
with barrier height $E^{\ddagger}$ (see Fig.~\ref{fig:schematic}b,c). Figure~\ref{fig:dissipation} shows numerical calculations (App.~\ref{app:simulation-details} presents details) of the steady-state excess power $\expow$, \quantity $\beta\TAFER$, and mechanical entropy production rate $\dot{\Sigma}^X$, as functions of the chemical driving strength $\Delta\mu$. For $N=3$ mechanical states, across all barrier heights the excess power is less than the entropy production rate, while for $N = 12$, the excess power is greater. Thus, even for the simple case of $N=3$ and $X$ tracking the current state of $Y$, the excess power $\expow$ can significantly differ from the entropy production rate. In contrast, the \quantity [Eq.~\eqref{transduced-power}] equals (as expected) the entropy production rate $\dot{\Sigma}^X$ by the mechanical system's dynamics, for all energy barriers $\beta E^{\ddagger}$ and numbers $N$ of mechanical states. 

\subsection{Excess power can become negative\label{sec:expowNeg}}
At steady state, the excess power equals the negative heat flow $-\beta\dot{\mc{Q}}^X$ due to $X$ dynamics, and is therefore not lower bounded by zero (see App.~\ref{app:excess-power-entropy-prod}). To illustrate this, we consider the periodic bistable potential (see Fig.~\ref{fig:schematic}d)
\begin{align}
    &\beta \epsilon_{xy}^{\rm PMF}=  \label{energy-pmf}  \\
    &-\ln\left[ e^{-\frac{1}{2}\beta E^{\ddagger}\cos 2\pi(x-y)} + e^{-\frac{1}{2}\beta E_2^{\ddagger}\cos 2\pi(x - y - \phi) + \beta\Delta E} \right] \nonumber \ .
\end{align}
the potential of mean force~\cite{frenkel_book} of two offset sinusoidal potentials. $\phi$ and $\Delta E$ represent, respectively, the relative phase shift and energy offset.  

\begin{figure}
    \centering\includegraphics[width=\columnwidth]{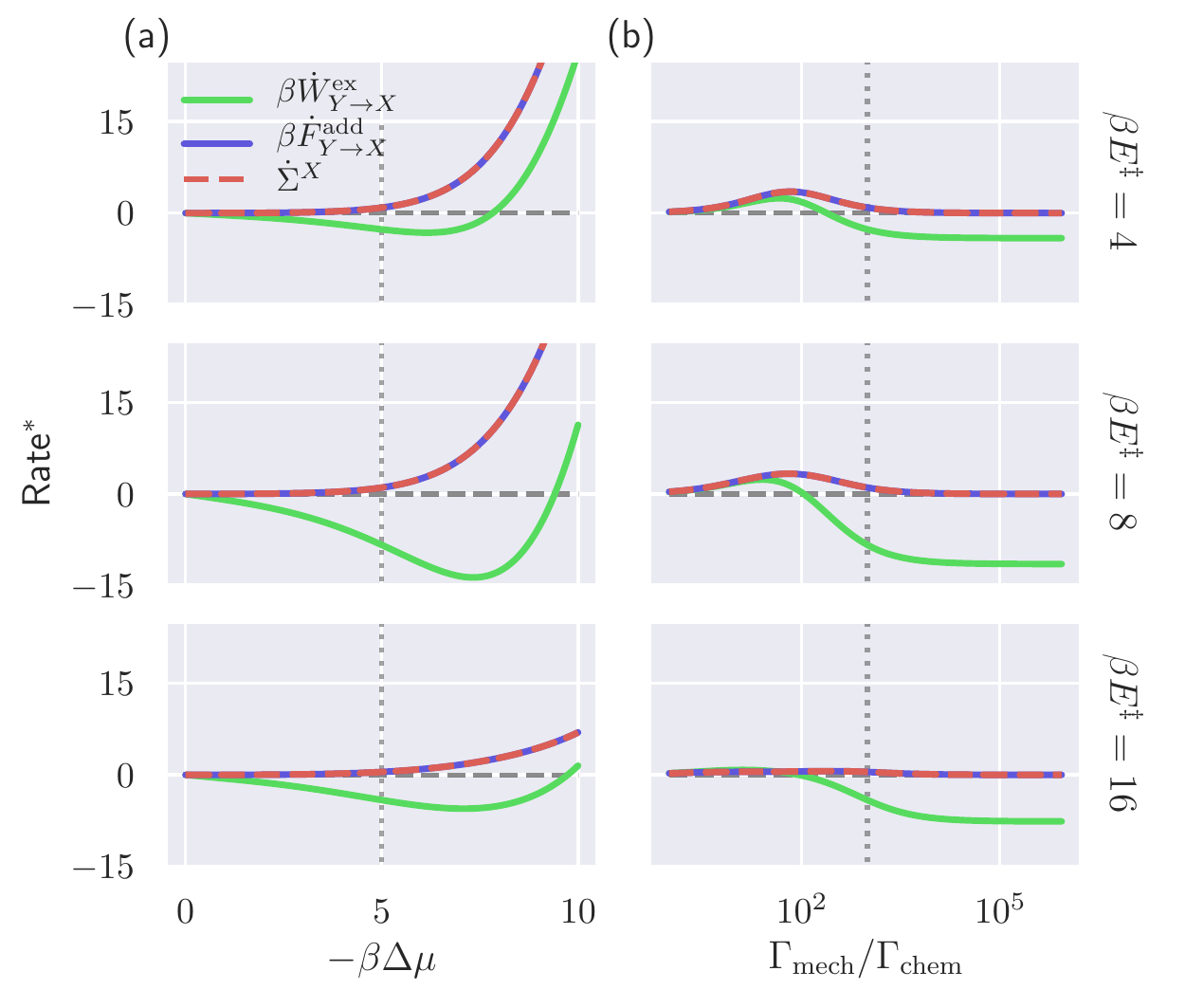}
    \caption{\label{fig:ex-power-neg}{\bf Steady-state excess power can become negative.} 
    Excess power $\expow$ (green), \quantity $\beta\TAFER$ (blue), and entropy production rate $\dot{\Sigma}^X$ (red dashed), as a function (a) of chemical potential difference $-\beta\Delta\mu$ at fixed ratio $\Gamma_{\rm mech}/\Gamma_{\rm chem} = 10^{4}$ of bare chemical and mechanical transition rates, and (b) of the ratio $\Gamma_{\rm mech}/\Gamma_{\rm chem}$ for fixed chemical potential difference $\beta\Delta\mu = 5$. Different rows show $\beta E^{\ddagger} = 4,8,16$. All calculations are at steady state for potential~\eqref{energy-pmf} and use $N = 12$, $E^{*} \equiv E^{\ddagger}/E_2^{\ddagger} = 1$, $\phi = 1/3$, and $\Delta E/E^{\ddagger} = 3/8$. Rates are nondimensionalized by the chemical bare rate: Rate$^*\equiv$ Rate$/\Gamma_{\rm chem}$. Dotted gray vertical lines indicate in (a) the chemical potential used in (b), and in (b) the bare rate ratio used in (a).}
\end{figure}

Figure~\ref{fig:ex-power-neg} shows the steady-state \quantity $\beta\TAFER$, excess power $\expow$, and entropy production rate $\dot{\Sigma}^X$ due to $X$ dynamics, as a function of (a) chemical potential difference and (b) the ratio $\Gamma_{\rm mech}/\Gamma_{\rm chem}$ of bare mechanical and chemical transition rates. Once again, TAFER equals the entropy production. The excess power is negative for intermediate chemical potential differences and for large rate ratios: $\Gamma_{\rm mech}/\Gamma_{\rm chem} \gtrapprox 10^{3}$.

Physically, the excess power drops below zero because the switching-state distribution $p_{yy'}^{{\rm sw}, x}$ for clockwise $Y$ transitions is dominated by mechanical states $x$ that lose energy during the transition. As a result, the excess power required to drive the system via $Y$ dynamics becomes negative. 

Thus even though the mathematical forms of the \quantity and the entropy production rate of the mechanical system are quite different---one involving a summation over $Y$ transitions at fixed mechanical state $x$, the other involving a summation over $X$ transitions at fixed chemical state $y$---they are equal at steady state. The excess power $\expow$ by itself is not lower bounded by zero and can become negative (Fig.~\ref{fig:ex-power-neg}). This reinforces the mathematical demonstration (Sec.~\ref{subsec:transduced-power}) that TAFER is the thermodynamic generalization of excess power to autonomous systems.

\section{\label{sec:discussion}Discussion}

Autonomous stochastic systems are models for many molecular machines, where strong coupling between multiple stochastic coordinates is central to their functioning. It is intuitively appealing to view the interacting components of a molecular machine as driving one another.

For (non-autonomous) systems where the driver--subsystem $Y$ in Sec.~\ref{sec:model-system}--receives no feedback from the driven subsystem ($X$), the excess power flowing from $Y$ to $X$ equals the entropy produced by the driven subsystem's dynamics, relating upstream energy flows (excess power) to downstream entropy production. It is tempting to make use of this correspondence between excess power and entropy production, but excess power does not play the same role in coupled, thermodynamically complete, autonomous systems and has no simple relationship with entropy production.

We introduced here the \quantity $\beta\TAFER$ as a measure of dissipation in strongly coupled stochastic systems that plays the same thermodynamic role as the excess power in systems driven by an external control parameter. In particular, $\beta\TAFER$ equals the steady-state entropy production rate during subsystem $X$'s dynamics, but is an explicit function of the dynamics of $Y$, which may be more easily observable in particular contexts~\cite{toyabe_2010,ariga_2018}. Furthermore, our investigation of the \quantity provides a convenient generalization beyond the reversible limit of the work associated with stochastic driving protocols~\cite{verley_2014}, and an analysis of the trade-offs between control work and mutual information complementary to Ref.~\cite{barato_2017}.

We expect that these insights will be useful for ongoing research in stochastic thermodynamics, extending theoretical results for systems driven by an external control parameter~\cite{large_2019,large_2018,sivak_2012,aurell_2011,schmiedl_2007} to autonomous models of molecular machines. A better understanding of dissipation in thermodynamically complete systems--where inter-system feedback satisfies local detailed balance--will clarify the functional capabilities and limitations of molecular machines. By consistently incorporating feedback into the control schema, we can further elucidate the rich physics in strongly coupled systems.

\acknowledgments
We thank Steven Blaber (SFU Physics) and Miranda Louwerse (SFU Chemistry) for insightful comments on the manuscript. 
This work is supported by Natural Sciences and Engineering Research Council of Canada (NSERC) Canada Graduate Scholarships--Masters and Doctoral (SJL), 
by Grant No. FQXi-IAF19-02 from the Foundational Questions Institute Fund, a donor-advised fund of the Silicon Valley Community Foundation (JE \& DAS),
an NSERC Discovery Grant (DAS), and a Tier-II Canada Research Chair (DAS).

%

\appendix

\section{\label{app:autonomous-vs-external}Entropy production in thermodynamically complete or incomplete systems}

Here we show how the entropy production of a subsystem differs under dynamics that are detailed balanced compared to dynamics that are not. In particular, for the model system in Sec.~\ref{sec:expowNEEP} we show that the entropy production rate only vanishes at zero driving when the system is thermodynamically complete. To start, we generalize the $Y$ transition rates,
\begin{equation}
    \W{yy'}{x} = \Gamma_{\rm chem}\exp\left\{-\tfrac{1}{2}\beta\left( \Delta\mu_{yy'} + \alpha\left[\Delta\epsilon_{yy'}^x\right] \right)\right\}
\end{equation}
where the feedback parameter $\alpha\in [0,1]$ interpolates the chemical dynamics between detailed balanced ($\alpha = 1$, Sec.~\ref{subsec:transduced-power}) and feedback-free ($\alpha = 0$, Sec~\ref{subsec:deterministic-work}). For all $\alpha \neq 1$ the system is thermodynamically incomplete, breaking local detailed balance~\eqref{local-detailed-balance}, and for decreasing $\alpha$ the feedback from the mechanical subsystem $X$ to the chemical dynamics decreases.

\begin{figure}[h]
\centering\includegraphics[width=\columnwidth]{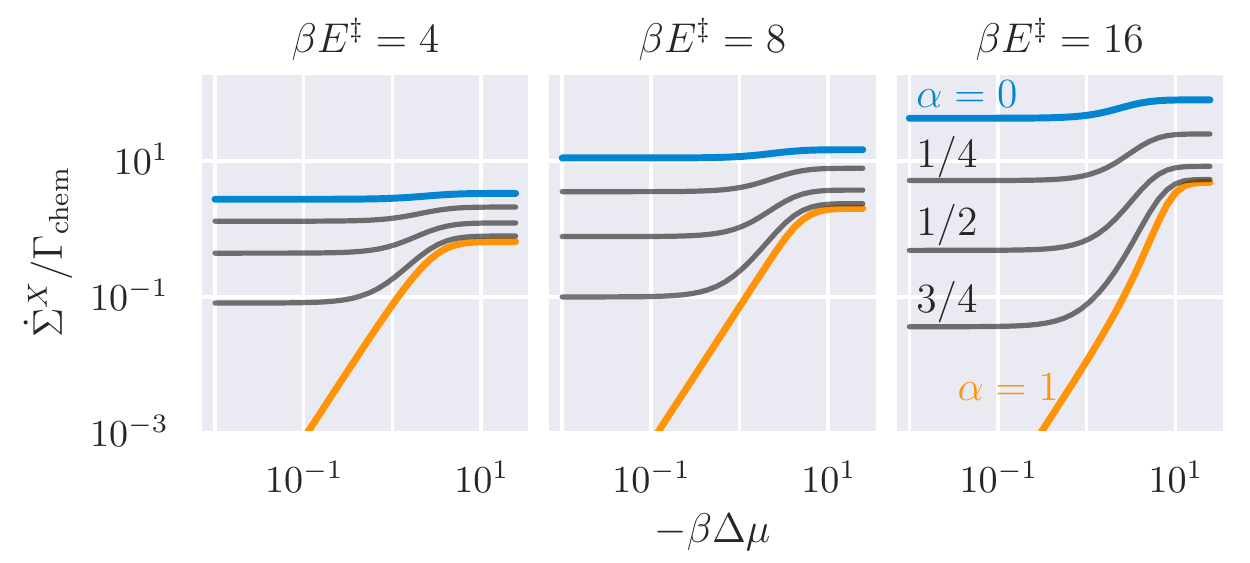}
\caption{\label{fig:suppfig_1}{\bf Dynamics that break detailed balance produce nonzero steady-state entropy production, even at $\boldsymbol{\beta\Delta\mu = 0}$.} Across all energy barriers $\beta E^{\ddagger} = 4,8,16$, only detailed-balanced joint dynamics ($\alpha=1$, orange curves) produce vanishing entropy production rate $\dot{\Sigma}^X$ as $-\beta\Delta\mu \to 0$. For all other $\alpha$ (blue and gray curves), the entropy production rate approaches a nonzero value as  $-\beta\Delta\mu\to 0$. $Y$ transitions that are independent of $X$ ($\alpha=0$, blue) have the greatest asymptotic value, while increasing feedback parameter $\alpha$ monotonically decreases the asymptotic entropy production rate. For all panels, the number of mechanical states is $N=12$, and the entropy production rate is nondimensionalized by the bare chemical transition rate: $\dot{\Sigma}^X/\Gamma_{\rm chem}$.}
\end{figure}

Figure~\ref{fig:suppfig_1} shows the entropy production rate due to subsystem $X$ dynamics as a function of the chemical driving strength $\beta\Delta\mu$, for energy barriers $\beta E^{\ddagger} = 4,8,16$. As $\beta\Delta\mu\to 0$, for a thermodynamically complete system ($\alpha=1$) the entropy production rate $\dot{\Sigma}^X$ vanishes, while for a thermodynamically incomplete system ($\alpha\neq 0$), $\dot{\Sigma}^X$ approaches a nonzero asymptotic value. The asymptotic value monotonically increases with decreasing $\alpha$, and is largest for $\alpha=0$, when $Y$ dynamics receive no feedback from $X$.

\section{\label{app:trans-power}Detailed derivation of \quantity}
At steady state, the rate of change of internal energy $E_{XY}$ is zero, thus
\begin{subequations}
\label{app-energy-rate}
\begin{align}
	0 &= \md_t E_{XY} \\
	&= \sum_{x,y,y'}\beta(\epsilon_{xy} - \epsilon_{xy'})\W{yy'}{x} p_{xy'} \\
	&\quad + \sum_{x,x',y}\beta(\epsilon_{xy}-\epsilon_{x'y})\W{y}{xx'} p_{x'y} \nonumber \ .
\end{align}
\end{subequations}
Substituting this into the entropy production due to $X$ dynamics~[Eq.~\eqref{entropy-split}], we separate the energetic contribution:
\begin{align}
    \dot{\Sigma}^X &= \beta \sum\limits_{x,y,y'} \W{yy'}{x} p_{xy'} \left( \epsilon_{xy}- \epsilon_{xy'} \right) - \beta \, \md_t E_{XY} \nonumber\\
    &\qquad+ \sum\limits_{x,x',y} \W{y}{x x'} p_{x'y}  \ln\frac{p_{x'|y}}{p_{x|y}}\ .
\end{align}
We identify the first RHS term as the power defined in Eq.~\eqref{autonomous-power}, while the second and third RHS terms can only be combined to a derivative of free energy if we include $\dot{I}^Y$, capturing the change in mutual information due to the $Y$ dynamics:
\begin{subequations}
\begin{align}
    \beta\, \md_t &E_{XY} - \sum\limits_{x,x',y} \W{y}{x x'} p_{x'y}  \ln\frac{p_{x'|y}}{p_{x|y}} \nonumber \\
    &= \beta\, \md_t E_{XY} - \sum\limits_{x,x',y,y'} \W{yy'}{x x'} p_{x'y'}\,\ln\frac{p_{x'|y'}}{p_{x|y}} \\
    &\qquad\qquad\qquad\qquad+\sum\limits_{x,y,y'} \W{yy'}{x} p_{xy'}  \ln\frac{p_{x|y'}}{p_{x|y}} \nonumber \\
    &= \beta\, \md_t F_{X|Y}^{\rm neq} - \dot{I}^Y\ .
\end{align}
\end{subequations}
Here, we used the definition [Eq.~\eqref{info-rate-y}] of the information rate due to $Y$ dynamics, and introduced the \emph{conditional nonequilibrium free energy}~\cite{Esposito2011NonequilibriumFreeEnergy},
\begin{align}
    \beta F_{X|Y}^{\rm neq} &\equiv \beta E_{XY} - S_{X|Y}\ . \label{app:neq-free-energy}
\end{align}
$E_{XY} \equiv \sum_{x,y} \epsilon_{xy}p_{x y}$ and conditional entropy $S_{X|Y} \equiv -\sum_{x,y}p_{x y}\ln p_{x|y}$ represent averages of the stochastic energy and conditional stochastic entropy, over the joint distribution $p_{x y}$. 

Consequently, we find:
\begin{align} \label{eq:excessPowerNoneq}
    \dot{\Sigma}^X = \beta \dot{W}_{Y\to X} - \beta\, \md_t F_{X|Y}^{\rm neq} + \dot{I}^Y \geq 0\ .
\end{align}
The \quantity is then defined using the conditional \emph{equilibrium} free energy $F_{X|Y}$ (the average over $Y$ of $F_{X|y}$) instead of its nonequilibrium counterpart:
\begin{align}
    \beta \TAFER = \beta \dot{W}_{Y\to X} - \beta\, \md_t F_{X|Y} + \dot{I}^Y\ . \label{app:transduced-power}
\end{align}

We can now more fully appreciate the similarity between \quantity and entropy production rate: For processes starting and ending in equilibrium, integrating the entropy production rate and the excess power give the same result. For steady-state systems, both measures agree because both the equilibrium and nonequilibrium free energy are unchanging. Furthermore, using the equilibrium free energy in Eq.~\eqref{app:transduced-power} results in an expression for the \quantity which is independent of $X$ dynamics, while the same is not true if the nonequilibrium free energy $F_{X|Y}^{\rm neq}$ [Eq.~\eqref{app:neq-free-energy}] is used, as its rate of change depends on the $X$ dynamics through the conditional entropy $S_{X|Y}$.

\section{\label{app:excess-power-entropy-prod}At steady state, excess power equals heat flow}

Here we derive---for detailed-balanced dynamics---the equality of the entropy production in the reservoir (heat flow) due to the dynamics of subsystem $X$, and the excess power done on subsystem $X$ by subsystem $Y$. We use Eq.~\eqref{app-energy-rate} to rewrite the excess power [Eq.~\eqref{autonomous-power}] as
\begin{subequations}
\begin{align}
	\expow &= \sum_{x,y,y'}\beta(\epsilon_{xy} - \epsilon_{xy'})\W{yy'}{x} p_{xy'} \label{app-power-entropy-2} \\
	&= -\sum_{x,x',y}\beta(\epsilon_{xy} - \epsilon_{x'y})\W{y}{xx'} p_{x'y} \label{app-power-entropy-3} \\
	&= \sum_{x,x',y}\W{y}{xx'} p_{x'y}\ln\frac{\W{y}{xx'}}{\W{y}{x'x}} \label{app-power-entropy-5} \\
	&= -\beta\dot{\mc{Q}}^X
	\label{app-power-entropy-6} \ .
\end{align}	\label{app-power-entropy}%
\end{subequations}
In Eq.~\eqref{app-power-entropy-5} we use local detailed balance of the microscopic rates~[Eq.~\eqref{local-detailed-balance}] along with the fact that there are no changes in chemical potential during $X$ transitions. In Eq.~\eqref{app-power-entropy-6} we substitute the definition of heat flow from Ref.~\cite{horowitz_2014}.

\section{\label{app:trans-power-add-free-energy}Transduced additional free energy rate and nonequilibrium free energy}

Substituting Eq.~\eqref{switching-state-dist} and Eq.~\eqref{boltzmann-conditional} into Eq.~\eqref{transduced-power-explicit}, the \quantity can be written as
\begin{align}
  \beta\TAFER &= \sum_{x,y,y'}\W{yy'}{x} p_{xy'} \ln\frac{\pi_{x|y'}p_{x|y}}{\pi_{x|y}p_{x|y'}} \\
  &= \sum_{x,y,y'}\W{yy'}{x} p_{xy'} \big[ \beta \left( \epsilon_{xy}- F_{X|y} - \epsilon_{xy'} +F_{X|y'} \right)\nonumber\\
  &\qquad\qquad + \ln p_{x|y} - \ln p_{x|y'} \big]\ . \label{eq:app_transducedWork}
\end{align}

We now define a \emph{stochastic nonequilibrium free energy}
\begin{align}
  \beta f_{x|y}^{\rm neq} \equiv \beta\epsilon_{xy} + \ln p_{x|y}\ ,  
\end{align}
which is the specific quantity that, when averaged over the joint distribution $p_{x y}$, gives the conditional nonequilibrium free energy~[Eq.~\eqref{app:neq-free-energy}].

This definition allows us to rewrite Eq.~\eqref{eq:app_transducedWork} as
\begin{align}
  \TAFER &= \sum_{x,y,y'}\W{yy'}{x} p_{xy'} \left[f_{x|y}^{\rm neq} -f_{x|y'}^{\rm neq} -\left( F_{X|y} - F_{X|y'}  \right) \right] \nonumber\\
  &= \dot{F}^{{\rm neq},Y}_{X|Y}- \mathrm{d}_t F_{X|Y} \ ,  
\end{align}
where $\dot{F}^{{\rm neq},Y}_{X|Y}$ is the change of conditional nonequilibrium free energy that is due to the $Y$ dynamics. Here, we utilized the splitting of rates of change introduced in Eq.~\eqref{time-deriv-decomp}. Finally, note that $X$ dynamics don't modify the average equilibrium free energy $F_{X|Y}$, thus $\mathrm{d}_t F_{X|Y}$ is 
due 
only 
to $Y$ dynamics.

\section{\label{app:simulation-details}Simulation details}
With rates given by Eqs.~\eqref{simulation-rates}, for a given (fixed) chemical driving $\Delta\mu$, we calculate the steady-state distribution by solving for the unique right eigenvector $p_{x'y'}^{\rm ss}$ corresponding to the zero eigenvalue of the rate matrix $\W{yy'}{xx'}$~\cite{van-kampen}:
\begin{equation}
    \W{yy'}{xx'} p_{x'y'}^{\rm ss} = 0 \ .
\end{equation}
We use the eigenvalue solver in the \texttt{scipy.linalg} python package to numerically calculate the steady-state distribution at each chemical driving strength. From this steady-state distribution, we directly calculate all the dissipation measures: $\dot{\Sigma}^X$, $\expow$, and $\beta\TAFER$.

\end{document}